\documentclass{aa}  

\usepackage{graphicx}
\usepackage{txfonts}
\newcommand{\kms}{km\,s$^{-1}$}
\newcommand{\lam}{$\lambda$}

\newcommand{\CaI}{\ion{Ca}{I}}

\newcommand{\LiI}{\ion{Li}{I}}

\begin{document}

   \title{Lithium in coalesced non-compact stars}
   \author{Tomek Kami\'nski\inst{1}
          \and Mirek Schmidt\inst{1}
          \and Marcin Hajduk \inst{2}
          \and Aleksandra Kiljan\inst{3}
          \and Inna Izviekova\inst{1}
          \and Adam Frankowski\inst{1}
          }
\institute{\centering
Nicolaus Copernicus Astronomical Center, Polish Academy of Sciences, Rabia{\'n}ska 8, 87-100 Toru{\'n}, Poland, \email{tomkam@ncac.torun.pl}\label{inst1}
\and Space Radio-Diagnostics Research Centre, University of Warmia and Mazury, ul. Oczapowskiego 2, 10-719 Olsztyn, Poland \label{inst2}
\and Warsaw University Astronomical Observatory, Al. Ujazdowskie 4, 00-478 Warszawa, Poland \label{inst3}
}
\authorrunning{T. Kami\'nski et al.}
\titlerunning{Li in merger products}

 
  \abstract
   {Galactic red novae are thought to be produced in stellar mergers between non-compact stars, such as main-sequence stars and cool giants. They are hoped to help in explaining physical processes involved in common envelope evolution and stellar binary collisions.}
   {We investigate the presence of lithium in three best-observed Galactic red nova remnants. Explaining the origin of lithium may point to mixing mechanism present before, during, or after the merger.}
   {The lithium line at 6707.81\,\AA\ was compared to a feature of [Ca\,I at 6572.78\,\AA\ to derive relative abundances in circumstellar gas. Absolute abundances were next calculated assuming the Solar calcium to lithium abundance ratio.}
   {Lithium abundances were measured in the merger remnants of V838\,Mon with $A$(Li)=2.3, CK\,Vul with $A$(Li)=2.5, and V1309\,Sco with $A$(Li)=1.8.}
   {Lithium is overabundant in red novae suggesting that at least some merger products activate the Cameron-Fowler mechanism whereby convective mixing can reach the deep stellar interior. Whether deep convection and associated diffusion alone or some other processes (e.g. spindown) can be responsible for driving the Cameron-Fowler mechanism in the remnants requires further studies. Early observations of lithium in V838\,Mon hint that these mechanisms activate early, perhaps already in the common envelope phase. These observations should be taken into account in modelling these complex systems.}


   \maketitle
%
\section{Introduction}
Red novae (or luminous red novae) are a group of eruptive stars that have been postulated to outburst as stellar mergers of non-compact stars \citep{TS2006, Tylenda2011, PastorelloRev}. We currently classify five Galactic objects in this group: CK\,Vul (a.k.a. Nova 1670), V4332\,Sgr (Nova Sgr 1994), V838\,Mon (Nova Mon 2002), V1309\,Sco (Nova Sco 2008), and EWS 2002-OGLE-BLG-360. While other eruptive stars, like supernovae and classical novae, red novae are primarily powered by accretion and, in consequence, do not reach high plasma temperatures. Their stellar remnants -- that is, the products of the coalescence -- are cool giant or supergiant like stars with effective temperatures of 2000--3000\,K. Owing to this low photospheric temperature and mass loss phenomena (at different stages), the circumstellar media of the merger products are remarkably rich in cool gas, often present in molecular form, and quickly become enshrouded in dust. The circumstellar envelopes of red nova remnants are very complex as they combine material lost prior, during, and after the merger. Many of the studied structures are asymmetric, pointing into formation of disks, jets, wide-angle outflows, and inhomogeneous winds. This complexity seems to hold the key to our understanding of physical mechanisms driving stellar systems to mergers and of the merger process itself.

Although still small in numbers, the Galactic red novae are rather an inhomogeneous group of variables. While V838\,Mon progenitor was a young ($<20$\,Myr, \citealp{AfsarBond}) triple system \citep{kamiV838alma} with an 8\,M$_{\sun}$ primary \citep{TylendaProgenitor}, most of the other objects were systems of solar mass stars which were evolved in the time of the outburst: V1309 Sco and V4332\,Sgr were contact systems of two subgiants \citep{Stepien}, while EWS 2002-BLG-360 and CK\,Vul were systems containing a red giant branch (RGB) star \citep{TylendaBLG, Kami26Al, MacLeod2022}. The timescales on which we have studied the remnants of Galactic red novae are typically short, up to decades (V4332\,Sgr erupted in 1994), but the outburst of CK\,Vul was observed in 1670--72 \citep{Hevelius1671,Shara}, giving us currently a timescale of over 350 years. The youngest Galactic red nova, V1309\,Sco, erupted in 2008. For a brief overview of the objects, see \cite{KamiSubmm}. 

Over a dozen extragalactic red nova analogs (a.k.a. luminous red novae or, more broadly, intermediate luminosity optical/red transients) has been identified in the Local Group \citep[e.g.,][]{BondM31RV,Pastorello2019,PastorelloRev,Pastorello1,Pastorello2,Blagorodnova2017,Blagorodnova2020,Blagorodnova2021,Williams,Williams2020,Kurtenkov,MacLeodM31,Smith2016,Rau,Cai}. However, information on their progenitors and remnants is sparse. Due to lack of adequate spectral observations, they are not the main subject of this study. 

In recent years, red novae have been recognized as objects important for understanding the physics of the common envelope evolution and stellar mergers. The focus on red novae especially intensified when the V1309\,Sco's progenitor, relatively well observed before the red nova outburst in 2008, was recognized to had been an eclipsing contact binary inspiraling into a common envelope phase \citep{Tylenda2011,Pejcha2014,Pejcha2016}. Observations of red novae and their remnants are used to investigate the mechanisms behind the common envelope evolution and merger events, including the associated mass loss phenomena \citep[e.g.,][]{Ivanova,Nandez,Soker2,Soker1,Soker3,Soker4,Soker5,Soker6,Soker7,MacLeod2022,MacLeod2018a,MacLeod2018b,MacleodLoeb,Pejcha2014,Pejcha2016,Pejcha2016b,PejchaMetzger,MetzgerPejcha,Iaconi2018,Iaconi2019,Reichardt}. 

On the course of detailed studies of individual red nova remnants, we noticed that at least three of them -- the three best observed objects, that is V838\,Mon, CK\,Vul and V1309\,Sco -- show clear signatures of lithium in their spectra. Since lithium is generally destroyed in stars, the spectral features usually imply lithium enhancement. Overabundance of lithium in stars, especially evolved ones, is not entirely understood but generally points to effective mixing mechanisms in stellar interiors \citep[e.g.,][]{LiMassanger}. Curious whether lithium in red novae can identify similar mixing mechanisms and by this shed light on the merger physics, we investigated the lithium features in the Galactic red novae using methods described in Sect.\,\ref{method}. We next constrained lithium content and discussed its origin in V838\,Mon, CK\,Vul, and V1309\,Sco in Sects.\,\ref{v838}, \ref{ck}, and \ref{V1309}, respectively. We briefly comment on the presence of lithium in V4332\,Sgr in a subsection of Sect.\,\ref{V1309}. The fifth known Galactic red nova, EWS 2002-BLG-360, is not discussed here as no relevant observational data exist for this heavily dust-enshrouded object \citep{TylendaBLG}. We draw our final conclusions in Sect.\,\ref{final}. 

\section{Methodology}\label{method}
We analyzed the presence of lithium in red nova remnants based on observations of a single feature of $^7$Li\,I at 6707\,\AA\ which is formed in the circumstellar gas, not in the photosphere. The feature is a tight doublet of lines separated by 0.15\,\AA\ or 6.7\,\kms. This split is insignificant compared to the local turbulence and kinematic broadening of the features in all analyzed objects. We assume a weighted mean central wavelength of 6707.81\,\AA\ as the reference wavelength of the doublet; we used the transitions' oscillator strengths as weights \citep{LiAtomic} in the calculation. Stars rich in lithium occasionally display also features near 6103.6 and 8126.5\,\AA\ which are non-resonant transitions and are not expected to be strong in cool circumstellar matter of our remnants. We do not find any photospheric Li\,I features at these wavelengths, either. Additionally, the $\lambda$6103 transition of Li\,I overlaps with a line of Ca\,I at 6102.72\,\AA\ and the information on the lithium content cannot be easily extracted from this spectral range. The 8126.5\,\AA\ transition is problematic because it is within a telluric band.

In order to derive relative and absolute (i.e. relative to H) abundances of Li\,I we compared the $\lambda6707$ feature to the resonance line of Ca\,I at 6572.78\,\AA. This Ca line is the only semi-forbidden transition among resonance lines observed toward our sources, and thus it is the least affected by saturation effects. Its profile is very similar to the Li\,I line profile in all our targets. Calcium features have been used in the analysis of lithium content in other objects \citep{WallersteinConti}, including classical novae \citep{izzo}. However, in circumstellar and interstellar matter, calcium is strongly depleted owing to condensation into dust, limiting its usefulness as a reference element. For instance, in the outflow of the carbon star CW\,Leo, calcium was found to be depleted by a factor of 300 with respect to the solar abundances \citep{MH2010} and by over 3000 in the diffuse interstellar medium \citep[ISM;][]{whittet}. Observed species with less significant depletion, e.g. Na or K, even though are observed in the same outflows of red nova remnants, have too saturated lines to be useful in our analysis. If calcium is depleted into dust in the analyzed outflows, our estimates on absolute abundances of lithium should be treated as upper limits. Depletion of lithium into dust is, in contrast, very small. For instance, \cite{ISMabundances} found it is depleted by a factor of 2 in the ISM.

Assuming both absorption lines are optically thin or have very similar optical depth, by measuring their equivalent widths, $W$, we can obtain the column density of Li\,I relative to Ca\,I from
\begin{equation}
\log \frac{N_{\rm LiI}}{N_{\rm CaI}}=\log \frac{W_{\rm Li} f_{\rm CaI}\lambda_{\rm CaI}^2}{W_{\rm Ca} f_{\rm LiI}\lambda_{\rm LiI}^2}.
\end{equation} 
Where $f$ and $\lambda$ are the oscillator strength and central wavelength of the transition, respectively. We adopted $f_{\rm LiI}$=0.75 for the Li\,I line, which is a sum of oscillator strengths of the doublet, and $f_{\rm CaI}=5.1\cdot10^{-5}$ for the Ca\,I line. Both species have similar ionization potentials (5.4 and 6.1\,eV for \LiI\ and \CaI, respectively) and thus we assume they have similar ionization fractions. We further assume that both species are chiefly in the ground state. This is justified by very low temperatures, of a few 100\,K, in the analyzed media. Then, the derived column density ratio can be interpreted as the relative abundance ratio. Adopting now the solar abundance of Ca of $A$(Ca)=6.36 \citep{SolarComposition} in our objects, we can constrain the Li abundance relative to hydrogen as $A({\rm Li})=\log(N_{\rm LiI}/N_{\rm CaI})+A({\rm Ca})$ (where $A({\rm X})=\log N_{\rm X}/ N_{\rm H}+12$).

\section{Lithium in outflows of V838\,Mon}\label{v838}
The \lam 6707 line of Li\,I has been one of the strongest atomic features in the post-outburst spectra of V838\,Mon (see below), but it was also clearly present during the outburst, which began at the beginning of January 2002 and had a visual light curve with three peaks. \cite{Munari2002} found a broad -- thus circumstellar -- feature of Li\,I as early as in late January, just before the main (second) outburst of V838\,Mon. In a quasi-photospheric spectrum observed in March 2002, i.e. shortly after the third outburst, \cite{KipperKlochkova} derived the lithium abundance of $A$(Li)=3.8. In an improved analysis of the same observational material, \cite{KipperConf} derived $A$(Li)=3.4. Thus, the lithium abundance was certainly super-solar ($A_{\sun}$(Li)=1.1\footnote{In both papers, the authors compared V838 Mon's lithium abundance to the meteoritic Solar System abundance, not to the photospheric values, and incorrectly concluded that V838\,Mon had a nearly solar lithium abundance.}, \citealp{SolarComposition}) and marginally super-meteoritic ($A_{\rm met}$(Li)=3.31) in March 2002. We take a closer look at the circumstellar lithium line monitored through high-resolution spectra obtained in much later epochs, from 2005 to 2020.

\begin{figure*}
    \sidecaption
    \includegraphics[trim=0 5 70 0, clip, width=12cm]{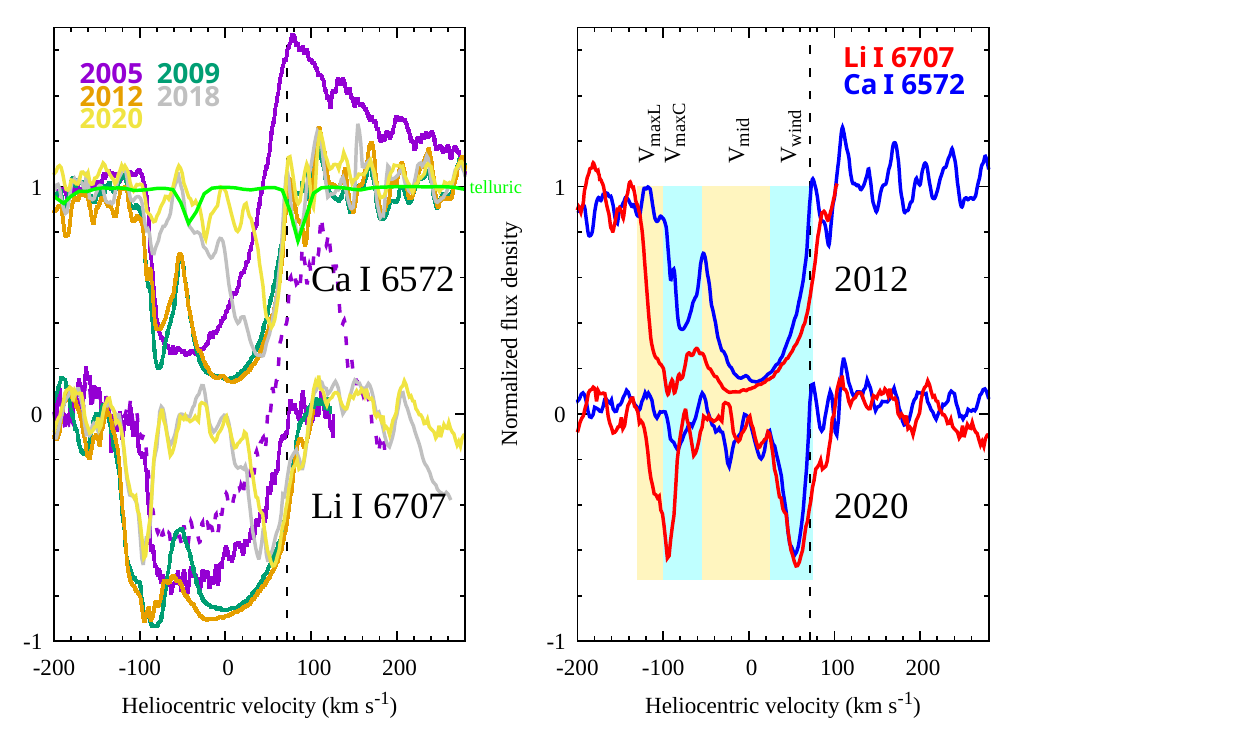}
    \caption{Profiles of the $\lambda$6572 line of Ca\,I and of the $\lambda$6707 line of Li\,I in V838 Mon. Normalization of the spectra is uncertain, in particular for the Li\,I profile from 2005 which may display a profile with (purple dashed) or without (purple solid) emission (see text). The alternative profile normalization for the 2005 epoch shown with dashed line was divided by 1.3 for better clarity. The Ca\,I profile is contaminated by two weak telluric features (light green), whose position and depth is different for all epochs, but is close to the sample telluric spectrum presented as a green line. The dashed vertical line indicates velocity of the SiO maser, which usually is indicative of a systemic velocity. The left panel shows changes in each profile over time, while the right panel compares profiles of both species for two sample epochs. Spectral regions highlighted in cyan and yellow were measured for relative line intensities (Table\,\ref{tab-measured-v838}).\\}
    \label{fig-V838-profiles}
\end{figure*}

The spectra used in our analysis represent five epochs and are listed in Table\,\ref{tab-V838-obs}. As illustrated in Fig.\,\ref{fig-V838-profiles}, the Ca\,I profile follows closely the variations in the Li\,I profile over the years. Both transitions, however, are located in spectral ranges where it is difficult to assess the local pseudo-continua. The presented profiles are normalized as a result of dividing the spectra by high-order polynomials, which were fitted in arbitrarily selected spectral ranges. The Li\,I line is located between the TiO $\gamma$ 1--0 $R_1$ and $R_2$ band-heads. Their shape influences the underlying ``baseline''. Depending on the actual shape of the TiO absorption bands, the Li\,I profile may possess or not an emission component at velocities higher than 80\,\kms. This is illustrated in Fig.\,\ref{fig-V838-profiles} for the 2005 epoch, for which we present two alternative normalizations of the Li\,I spectrum. This epoch was the most challenging case. 

The Ca\,I line, on the other hand, is heavily polluted by overlapping absorption bands of the TiO$\gamma$ (5--3, 6--4) and TiO$\gamma^{\prime}$ (0--1) systems, which makes a correction for the baseline difficult, too. For both species, we attempted a normalization through division by spectra of other cool stars or by TiO simulations, but the temperature and velocity structure of the TiO circumstellar gas of V838\,Mon is too complex for this approach to work satisfactorily well (but see an example in Fig.\,\ref{fig-v838-fit}). We therefore performed further analysis with spectra normalized by polynomials. In this approach, we assume that the atomic line profile is formed by absorbing from the TiO absorption spectrum, which may not be the case if TiO and neutral atomic gas are well mixed along the line of sight. Full radiative transfer including all opacity sources would be necessary to make a more realistic analysis. Also, all profiles are contaminated by low-amplitude TiO features, on the order of 0.1 of continuum, and Ca\,I line is additionally  affected by weak telluric features. All these shortcomings increase uncertainty in the derived abundances, but do not change our overall conclusions. 

\begin{table*}[]
    \centering
    \caption{High-resolution observations of V838\,Mon used in the analysis of the Li\,I line.}
    \begin{tabular}{cccccc}
    \hline
    Epoch & Obs. date (UT) & Telescope & Instrument & Resolution, $R$ & Reference \\
    \hline
    2005 & 13.10.2005  & Keck & HIRES& 34\,000 & 1\\
    2009 & 02.01, 13.02, 22.03.2009  & VLT  & UVES & 60\,000 & 2 \\
    2012 & 06.01.2012  & VLT  & UVES & 12\,000 & 3\\
    2018 & 27.10.2018  & Subaru &HDS & 20\,000 & 4 \\ 
    2020 & 23.12.2020  & SALT & HRS  & 40\,000 & 5 \\ 
    \hline
    \hline
    \end{tabular}
    \tablebib{(1) \cite{KamiKeck}, (2) \cite{Tylenda2011v838}, (3) \cite{KamiSubmm}, (4) Kami\'nski, in prep., (5) \cite{kamiV838alma} }
    \label{tab-V838-obs}
\end{table*}

The profile evolution of both lines is very similar. The P-Cyg profile from 2005 changed into a double broad absorption-dominated feature in 2009. This slight doubling had continued to 2012, but some time after that, the two subcomponents had split entirely, so that since $\approx$2018 they have been two well separated velocity components. Both components in both lines become weaker with time, but the Ca\,I component at most negative velocities is barely distinguishable from baseline variations in 2020. At a closer look, the profiles of both species display, however, some differences. A Li\,I component at extreme heliocentric velocities, i.e. $\le -95$\,\kms, has no corresponding absorption in the Ca\,I profile. Hereafter, we call this Li-rich and Ca-poor component $V_{\rm maxL}$. A component, which formed the well separated absorption in Ca\,I in later epochs, between about --100 and --55\,\kms, is designated $V_{\rm maxC}$. The range of intermediate velocities, between about --55 and +25\,\kms, where absorption nearly disappeared in later epochs, is hereafter designated as $V_{\rm mid}$. All the components mentioned so far can be assigned to outflows caused by the 2002 eruption that is known to had been associated with mass loss at changing terminal velocities. In the velocity range between about 25 and 75\,\kms, the latter being the systemic velocity measured from SiO maser observations \citep{Deguchi2005,Ortiz-Leon}, in late epochs we observe a well separated component which we assign to the ongoing mass loss (wind) from the coalesced star \citep[cf.][]{TylendaSpec,kamiV838alma}. We designated it $V_{\rm wind}$. All components are graphically presented in Fig.\,\ref{fig-V838-profiles}. There may be an extra narrow component at velocities higher than the maser, indicative of a fallback onto the star, but it is too weak to be analyzed here.

\begin{table}[]
    \centering
    \caption{Equivalent widths of the Li\,I and Ca\,I lines in V838\,Mon.}
    \label{tab-measured-v838}
    \begin{tabular}{ccc ccc}
    \hline\hline
         &\multicolumn{5}{c}{$W$ (\AA) for Li\,I}\\
    Component&	2005\tablefootmark{a}&	2009&	2012&	2018&	2020\\
\hline
total\tablefootmark{b}  &	1.8	&3.4	&3.6	&1.1	&1.1\\
$V_{\rm maxL}$&		    &0.4	&0.4	&0.2	&0.2\\
$V_{\rm maxC}$&		    &0.8	&0.8	&0.3	&0.2\\
$V_{\rm mid}$&		    &1.4	&1.5	&<0.1	&<0.1\\
$V_{\rm wind}$&		    &0.8	&0.8	&0.6	&0.6\\
\hline
         &\multicolumn{5}{c}{$W$ (\AA) for Ca\,I}\\			
    Component&	2005&	2009&	2012&	2018&	2020\\
\hline
total\tablefootmark{b}  &	1.5	&2.4&	2.4&	0.8&	0.6\\
$V_{\rm maxL}$&		&<0.1&	<0.1&	<0.1&	<0.1\\
$V_{\rm maxC}$&		&0.5&	0.5&	<0.1&	<0.1\\
$V_{\rm mid}$&		&1.3&	1.3&	0.3&	0.2\\
$V_{\rm wind}$&		&0.5&	0.6&	0.47&	0.4\\
\hline					
	&\multicolumn{5}{c}{log of the Li to Ca abundance}\\			
    Component&	2005&	2009&	2012&	2018&	2020\\
\hline
total\tablefootmark{b}&	--4.1&	--4.0&	--4.0&	--4.0&	--3.9\\
$V_{\rm maxL}$&	&&&&				\\
$V_{\rm maxC}$&&		--4.0&	--4.0&&		\\
$V_{\rm mid}$&&		--4.2&	--4.1&&		\\
$V_{\rm wind}$& &--4.0	&--4.1	&--4.0	&--4.0\\
\hline\hline
    \end{tabular}
    \tablefoot{\tablefoottext{a}{Spectrum was normalized as showed with the dashed line in Fig.\,\ref{fig-V838-profiles}.}\tablefoottext{b}{Intergration over the entire profile.}}
\end{table}

We measured equivalent widths of the different components and derived the relative and absolute abundances of Li\,I in V838\,Mon. The results are presented in Table\,\ref{tab-measured-v838}. The ratios we obtain for different epochs and different parts of the outflow are very consistent and on average give $\log(N_{\rm LiI}/N_{\rm CaI})$=--4.0. This constrains an absolute abundance of lithium at $A({\rm Li})$=2.3, an order of magnitude lower than that constrained from 2002 quasi-photospheric spectra. At the same time, in the component $V_{\rm maxL}$ at the highest outflow velocity, where no Ca\,I emission is observed, the lithium abundance should be at least one order of magnitude higher, i.e. close to or higher than the meteoritic value. 

We have assumed that the analyzed lines are optically thin or at least have a similar optical depth. By modelling of the Li\,I P\,Cyg features observed in V838\,Mon in 2005 using the same techniques as in  \cite{TylendaSpec} (their Sect. 5.2), we found a mean optical depth of the \lam 6572 line of $\tau \la 4$ --- the line was only moderately thick. The lines very likely became even optically thinner in later epochs due to expansion. 

\subsection{Origin of lithium in V838\,Mon}
The presence of lithium in the remnant of V838\,Mon is especially perplexing when one realizes that before 2002 the primary of the coalesced system was a B3 main-sequences star of a mass of 8\,M$_{\sun}$ and with a radiative envelope. There should be practically no lithium in the primary. The observations show, however, that lithium was present in the early phases of the outburst, possibly at very high, nearly meteoritic, abundances. Our observations show that lithium was indeed enhanced in the matter lost during the 2002 event, and is present at a similar abundance in the wind of the coalesced star, which indirectly indicates lithium is also present in the photosphere. We consider generally two mechanisms that could be responsible for lithium in V838\,Mon: ({\it i}) internal lithium production in the common envelope system or in the coalesced star and ({\it ii}) an external enrichment through the material of the lower-mass star that merged with the B-type primary in 2002. 

\paragraph{Internal lithium enrichment}
The internal production scenario is justified by the similarity of the post-outburst remnant of V838 Mon to stars on the asymptotic giant branch \citep[AGB;][]{Sokermagnetic} and to red supergiants. The coalesced star mimics well a red giant or a red supergiant in the sense that it has a large luminosity of 10$^4$\,L$_{\sun}$ and huge dimensions with a 460\,R$_{\sun}$ radius \citep{kamiV838alma,Olivier}, and has a convective envelope that puts it close to the Hayashi limit on the Hertzsprung-Russel diagram. Lithium, or more specifically $^7$Li, can be produced in some RGB and AGB stars through fast mixing in the Cameron-Fowler mechanism (a.k.a. the ``$^7$Be-mechanism") \citep{Cameron1955, CameronFowler1971, Sackmann}. In this process, $^7$Be is produced in the second proton-proton chain (pp II), via $^3{\rm He}(\alpha,\gamma)^7{\rm Be}$, in deep internal layers of the star. In the complete pp II cycle, beryllium is next transformed into $^7$Li, which in turn is quickly destroyed in a reaction with a proton, $^7{\rm Be}({\rm e}^-,\nu)^7{\rm Li}(p,\alpha)^4{\rm He}$. However, fast mixing is thought to be able to effectively transfer the freshly produced $^7$Be to outer cooler layers where it decays into $^7$Li in conditions where lithium is not entirely destroyed and can be farther dredged up to the surface. Convection seems to be the major mixing mechanism explaining surface $^7$Li abundances in some AGB stars (including hot bottom burning), while in RGB stars an extra mixing mechanism is necessary, be it termohaline mixing, magneto-hydrodynamic instabilities, or rapid rotation, to mention just a few of the proposed mixing mechanisms \citep{Book,LiMassanger}. The currently measured lithium abundance of 2.3 in V838\,Mon is very close to average values of 2.3 and 2.6 measured in RGB and AGB giants \citep{MacielCosta2012}. We propose that the 2002 event turned the main-sequence star of 8\,M\,$_{\sun}$, i.e. a star with a radiative envelope, into a giant/supergiant with a fully convective envelope that can reach deep enough to dredge up $^7$Be and produce $^7$Li in the photosphere. This way, V838\,Mon would be a more extreme case of a scenario considered by \cite{Denissenkov2000} in which a planet or brown-dwarf engulfment can trigger an active Cameron-Fowler mechanism; in the case of V838\,Mon a much heavier body was accreted, making this process even more effective. 

Given that high lithium abundances were already observed in V838\,Mon during the 2002 outburst, the hypothesized $^7$Li production mechanism must have been activated very early on. It is tempting to assign this behavior to the common envelope phase, which the system went through just before the coalescence. Whether the presence of lithium enrichment is related to V838\,Mon being a merger in addition to being a common event is an open question which would require dedicated modeling. We note, however, that convective mixing that does not reach the H-burning regions would actually decrease the abundance of Li in the photosphere (by dredging up Li-poor material from internal layers where lithium is easily destroyed by protons). 

Alternatively, rotation of the remnant may provide extra mixing that can support the Cameron-Fowler mechanism. Some stellar merger models predict that the collision product contains a lot of angular momentum and thus is rapidly rotating \citep[e.g.][]{Schneider2019}. There is, as of now, no direct indication of a fast rotation of V838\,Mon, with an upper limit of $\varv \sin i < 28$\,\kms), but if this is ever found to be the case, internal rotation might provide extra mixing. Moreover, since such a star is also supposed to spin down with time, extra mixing is even more likely, but perhaps operates on longer time scales than the 20\,yr life-span of the remnant of V838\,Mon.


\paragraph{External lithium enrichment}
In the V838\,Mon merger scenario of \cite{TS2006}, the main sequence B-type star (or an A-type protostar) collides with a low-mass protostar of a mass of 0.1--0.5\,M$_{\sun}$. The latter mass constraint is very uncertain, since it was calculated as a mass needed to be accreted on the 8\,M$_{\sun}$ star to account for the observed power of the outburst. The young age of the progenitor is implied by its membership in a small open cluster \citep{AfsarBond}, which is still surrounded by large quantities of molecular gas and dust \citep{KamiEcho,TylendaKaminskiEcho}. The presence of early B-type main-sequence stars implies an upper limit on the age of the cluster of $\le$25\,Myr, but observations of the translucent molecular cloud near the cluster suggest an even younger age, 3--10 Myr. The youngest low-mass protostars are known to be lithium-rich ($A ({\rm Li})) \approx 3$) but, as they are fully convective, they lose lithium very quickly during their evolution to the main sequence after reaching core temperatures above about 3\,MK. It is estimated that the Li depletion timescale is the quickest for stars of a mass of 0.6\,M$_{\sun}$, which deplete lithium in about 20\,Myr \citep{WhiteHillenbrand2005}. The time scales are comparably quick for masses considered for the accreted star in V838\,Mon \citep{Bildsten1997}. For instance, for a 0.3\,M$_{\sun}$ protostar, 100-fold Li depletion is expected to take 28\,Myr and a lithium reduction by a factor of 2 takes 18\,Myr \citep{Chabrier1996}. 

If the age of the V838\,Mon progenitor is at the lower end of the suggested range, i.e. $<$10\,Myr, the low mass star could still have been rich in Li. This lithium-rich material could have been seen during the outburst and is seen now in the wind (and thus photosphere) of V838 Mon. Stellar-merger simulations show that the matter of the lower-mass body forms the outer layers of the merger product \citep[e.g.][]{TS2006,Sills}, which would explain the persistent presence of Li in the remnant. Ongoing convection and other mixing mechanisms that do not tap the hot $^3$He-rich layers will decrease the Li abundance with time. Additionally, the low-mass protostar could have been surrounded by a protoplanetary disk or other form of circumstellar matter which is expected to be Li rich, as it would be enhanced in lithium produced by spallation. We calculate that to produce the observed lithium enhancement in an envelope of 0.1\,M$_{\sun}$, the primary would have to accrete 0.01\,M$_{\sun}$ or 10 Jupiter masses of material with meteoritic abundance of lithium. This is not impossible given typical masses of protoplanetary disks of low-mass stars \citep[e.g][]{perseus}.

If the progenitor age is closer to 25\,Myr or the actual lithium depletion is more effective in low-mass stars than described by theoretical models, the internal production of lithium remains the more likely mechanism responsible for the observed lithium in V838\,Mon. Some observations indicate that low-mass protostars (as young as 8\,Myr) may be Li depleted sooner than theoretically expected \citep{WhiteHillenbrand2005, Yee2010, Jeffries2017}. Because lithium enhancement is seen in other remnants of red novae, we favor the scenario where lithium is commonly produced by coalesced stars internally.

The proposed scenarios result in lithium abundances with different $^7$Li/$^6$Li isotopic ratios. While the external enhancement should result in the cosmic $^7$Li/$^6$Li ratio of about 7, the internal production of lithium should considerably increase the content of the $^7$Li isotope only. Unfortunately, deriving the lithium isotopic ratio is very difficult even for sharp interstellar lines, and it is hopelessly hard, if not impossible, for the wind features in V838\,Mon. The internal production of lithium predicts also anomalous abundance of Be and B, but lines of those species are hardly available for observations. 

\subsection{Li-bearing molecules}
With the premise that lithium may be very highly enhanced in the circumstellar medium of V838\,Mon, we investigated whether there may be an enhanced presence of Li-bearing molecules in the remnant. A detection of such molecules would be instrumental in determining the dominant isotope of lithium. Overall, observations of Li-bearing species in astronomical environments are very rare \citep{LiHnondet}. We considered two species, LiOH and LiCl, which are thought to be among the main Li-bearing species that are easily accessible for millimeter-wave spectroscopy through their pure rotational transitions. We used the {\tt GGchem} code \citep{GGchem} to investigate molecular abundances in gas at different temperatures and pressures, assuming the thermochemical equilibrium. The latter assumption is questionable for the circumstellar gas of the red nova remnant, but we used the calculation only as a first guideline. We find that even if lithium abundance were extremely high, with $A({\rm Li})$=3--5, the Li-bearing molecules would have low absolute abundances and would produce lines that are not easily detectable with modern submillimeter interferometers like ALMA. In particular, LiOH emission should be 1--3 orders of magnitude weaker than AlOH lines detected in V838\,Mon at a modest S/N \citep{kamiV838alma}.

\section{Lithium in the CK\,Vul outflow}\label{ck}
The photosphere of the stellar remnant of Nova 1670 is not observable, as it is deeply embedded in circumstellar dust. The presumed merger product is surrounded by dusty bipolar nebula, which contains cool molecular gas as well as shock-excited atomic and H$_2$ matter \citep{KamiCKalma}. \citet{HajdukVarStars} noticed two variable stars towards the southern lobe of that nebula. The variability can be only explained if these are far background stars affected by changing extinction originating from the rapidly expanding southern lobe of CK\,Vul. In an optical spectrum of the brightest of the variable stars, Hajduk et al. found a strong lithium feature near 6707\,\AA\ which most likely originates in the outflow of CK Vul. Here we take a closer look at this feature, aiming to constrain the abundance and origin of lithium in the circumstellar medium of CK\,Vul.

We used a spectrum of \citet{HajdukVarStars} obtained with the Gemini Multiobject Spectrograph (GMOS) in 2010 to constrain first the parameters of the background star. We assume that the spectrum is correctly calibrated in fluxes, and we made no extra corrections for chromatic slit losses or effects caused by the atmospheric chromatic aberration. Using a grid of synthetic spectra of AMBRE \citep{ambre}, we found that the spectrum is best approximated by a photosphere of an effective temperature of 4750$\pm250$\,K (near spectral type $\sim$K3) and reddening (interstellar plus circumstellar) of $A_V$=6.0\,mag. Results of the best-fit search are shown in Fig.\,\ref{fig-ck-fit}. We used the standard interstellar extinction curve \citep{ccm} with the selective extinction parameter of $R_V$=3.1. Since no uncontaminated photospheric features are present, the luminosity class of the photosphere remains unconstrained. The effective temperature is rather well constrained. Cooler stars would produce a conspicuous MgH band near 5200\,\AA, which is not present, and hotter stars would be betrayed by a H$\alpha$ line which would be much stronger than the observed one. The derived temperature is close to that of a spectral type K4 ($T_{\rm eff}\!\approx$4500\,K) assigned to the star by Hajduk et al. However, the $V$-band reddening of 6.0 mag we derived is significantly higher than 4.4 mag postulated earlier. 

\begin{figure*}
    \sidecaption
    \includegraphics[trim=0 10 0 0,clip, width=12cm]{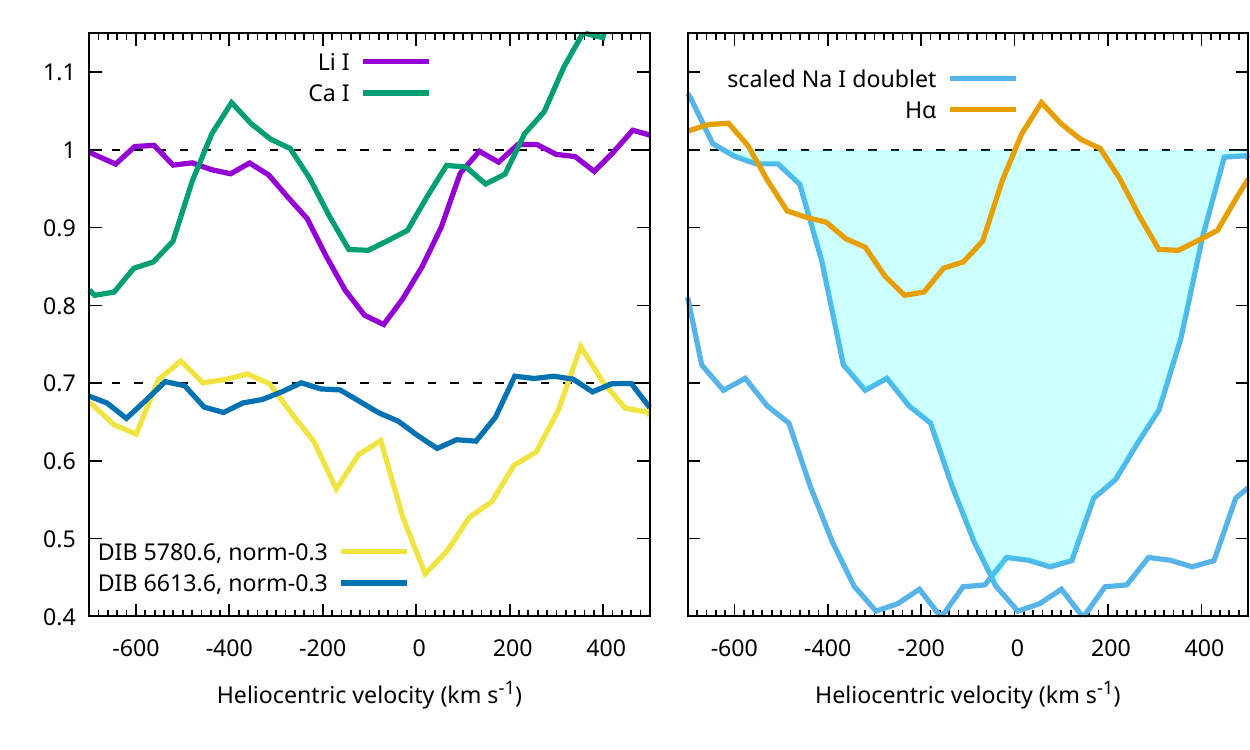}
    \caption{Profiles of absorption lines towards the star located behind the southern lobe of the CK\,Vul outflow. Left panel shows circumstellar absorption of Li\,I and Ca\,I in the circumstellar gas (upper) and interstellar features (shifted down by 0.3). Note different central velocities of the two groups. The right panel shows features of the Na\,I doublet and of H$\alpha$ which combine circumstellar, interstellar, and photospheric absorption. The filled cyan area shows absorption common for both lines of the blending Na\,I doublet and approximates the shape of the intrinsic Na\,I absorption.\\}
    \label{fig-ckvul-profiles}
\end{figure*}

Hajduk et al. measured the $R$ magnitude of the background star to be lower than 19\,mag in 2010. Adopting a spectral type K3 and assuming the star is on the main-sequence, we would expect $(V-R)$ of 0.8 mag and then the dereddened $V$ magnitude would be 18.6. Assuming now an absolute magnitude of 6.9\,mag for a K3\,V star, we obtain a distance of 2.2\,kpc. {\it Gaia} parallax measurements of the background star \citep[0.0660$\pm$0.1858\,mas][]{gaia} imply a 3$\sigma$ lower limit of 2.6\,kpc for the distance, so a luminosity class V can still be reconciled with the {\it Gaia} limits. However, higher luminosity classes, like giants, are more likely because the distance to CK\,Vul is much larger than 3\,kpc \citep{KamiCKdistance,BanerjeeCK}. Since only a small fraction, 1--2\%, of field giants have strong lithium features in their spectra \citep[e.g.][]{Smiljanic}, it is very unlikely we observe one here.  
 
There are no clear photospheric features in the spectrum of the background star, but, as mentioned, Na\,I and H$\alpha$ seem to have a photospheric component partially blending with interstellar and outflow absorption components. By comparing profiles of the Na\,I lines and H$\alpha$ to circumstellar and interstellar lines, we tentatively assign the photosphere a heliocentric velocity of $\approx$--250$\pm$50\,\kms. This is a much more negative velocity than the dominant interstellar and circumstellar components (cf. Fig.\,\ref{fig-ckvul-profiles}). 

The spectrum contains pure interstellar features. Hajduk et al. mentioned that the strongest diffuse interstellar band (DIB) is near 6284\,\AA, but we find it too close to a telluric band. Instead, the strongest DIB we have identified is the one near 5780.59\,\AA. There is also possibly a DIB at 6613.56\,\AA.

The most interesting for the discussion here are the circumstellar components. The lithium line $\lambda$6707 is clear from any contamination -- the closest features known in this source from other observations are the $\lambda$6716 line of the [\ion{S}{II}] doublet and $\lambda$6678 of He\,I,  which, fortunately, do not overlap with the Li line. Given the virtual absence of other photospheric lines, the line of lithium is very unlikely to be contaminated by absorption intrinsic to the field star, even if it happened to be a Li-rich giant. The Ca\,I line, when observed at the modest resolution of the GMOS spectra, is partially contaminated by an H$\alpha$ feature, which itself combines photospheric absorption and absorption from the CK\,Vul lobe (at different velocities; possibly also weak emission may be present). Despite this complication and as shown in Fig.\,\ref{fig-ckvul-profiles}, the velocity profiles of the Li\,I and Ca\,I lines overlap closely in velocity. This suggests that they are formed in the same region with a central heliocentric radial velocity of about --110\,\kms\ and the lines can be directly compared. Their radial velocity is certainly different from that of the interstellar gas probed by DIBs, which we measure is centered near 40\,\kms. They also overlap with the velocity range between --180 and 0\,\kms\ occupied by the undoubtedly circumstellar CO $J$=1--0 emission observed by ALMA along the same line of sight. 


The equivalent widths of $W_{\rm LiI}$=1.3 and $W_{\rm CaI}$=0.6 (both $\pm$0.1) \AA\ yield $\log(N_{\rm LiI}/N_{\rm CaI})$=--3.8 ($\pm$0.1). In a standard solar composition, this implies a lithium abundance of $A$(Li)=2.5. This lithium abundance is by 1.4 dex higher than in the Sun, but sill well below the meteoritic value. 

\subsection{Interpretation of Li in CK Vul}
Formation of Li in CK\,Vul may be different from that in V838\,Mon. There is a large body of evidence that CK\,Vul's circumstellar material has been heavily processed by nucleosynthesis processes associated with the progenitor and the merger event. Observations of optical atomic lines of the inner nebular remnant indicate a highly processed elemental composition. Foremost, helium is approximately twice more abundant relative to hydrogen than in the Solar photosphere and the N/O ratio is nearly ten times higher than in the Sun \citep{TylXshoot}. Millimeter-wave observations of cool molecular gas of the remnant of CK\,Vul also suggest a general overabundance of nitrogen \citep{KamiNat}. They additionally yield numerous isotopic ratios, for instance $^{12}$C/$^{13}$C$\approx$3.8, $^{14}$N/$^{15}$N$\approx$20, and $^{16}$O/$^{18}$O$\approx$36 \citep{KaminSingleDish} which together are very unusual. The chemical characteristics can be coherently explained by processing in the CNO hydrogen-burning cycles and by incomplete helium burning. Moreover, based on an unprecedented detection of an AlF isotopologue containing $^{26}$Al, which implies $^{27}$Al/$^{26}$Al$\approx$7, it was postulated that the progenitor system included an RGB star \citep{Kami26Al}. (Currently, the most likely scenario is a merger of an RGB star with a He white dwarf; Tylenda et al., in prep.). The analysis of optical spectra of heavier elements (Ne, S, and Ar) show additionally that the progenitor had a low metallicity with [Fe/H]=--0.53$\pm$0.07 \citep{TylXshoot}. While the general nucleosynthesis channels responsible for the peculiar composition can be identified, their relation to the stellar nucleosynthesis, nuclides processing during the presumed merger, or how exactly the matter was dispersed are currently not understood.   

The level of overabundance of Li in CK\,Vul is close to these observed in other Li-enriched objects, including RGB and AGB giants (J-type stars among them), classical novae, or R\,CrB (RCB) stars. Relevant for the discussion here is especially the similarity to RCB stars and J-type carbon stars, which are thought to be products of mergers between either CO and He white dwarfs (RCB stars), or He white dwarfs and red giants (J stars). The merger of CK\,Vul was certainly different from that postulated for RCB stars, chiefly due to the presence of a hydrogen envelope, but models of the latter still offer an informative analogy. In nucleosynthesis models of merging white dwarfs (WDs), $^7$Li is produced from $^3$He through $^7$Be, i.e. $^3{\rm He}(\alpha,\gamma)^7{\rm Be}({\rm e}^-,\nu)^7{\rm Li}$, just as in the Cameron-Fowler mechanism. White dwarf mergers are thought to effectively produce $^7$Be and thus $^7$Li only if $^3$He is abundantly available from earlier processing or directly from the matter of the He WD. It is also expected that the Li-rich region is mainly located in the outer layers of the merger product \citep{Longland}, which gives a chance of identification of the species in stellar spectra. 

Models of mergers leading to the formation of J (and early-R) type carbon stars -- that is, those that involve a red giant and thus are closer analogs of CK\,Vul -- are similar to RCB formation models. They produce lithium only provided that large quantities of $^3$He are mixed down to the He-burning layer and that freshly produced lithium has a way to escape to the outer layers of the coalesced star  \citep{Jstars2013,Jstars2020}. In those models, lithium abundances steeply increase over the first $\sim$1000\,yr after the merger and can reach super-meteoritic values. The lithium abundance of 2.5 dex in CK\,Vul's outflow observed some 350\,yr after the merger can be reconciled with these models, but more studies on the kinematic age of the Li-bearing outflow are necessary to assure this, as there is chance that the outflow is younger than 350 yr \citep{KamiCKdistance}. As mentioned, observations of CK\,Vul indicate an abundance of He that amounts to over half the mass of the optical nebula. There was therefore enough raw material to produce Li in the $^3{\rm He}(\alpha,\nu)^7{\rm Li}$ process. 

Dedicated merger models are necessary to verify the origin of Li in the extended bipolar remnant of CK\,Vul.



\section{Lithium in the emission spectrum of V1309\,Sco}\label{V1309}
V1309\,Sco is an important member of the red nova group, as it is the only object with well documented photometric variability prior to the merger in 2008 \citep{Tylenda2011}. Located in a dense field, it is a challenging source for optical observations. After around 2010, its spectral evolution could have been traced mainly through emission lines with no direct observations of the stellar continuum, which only becomes dominant in the mid-infrared \citep{Kamiv1309,KamiSubmm}.

\begin{figure*}
    \sidecaption
    \includegraphics[trim=25 285 20 0, clip, width=12cm]{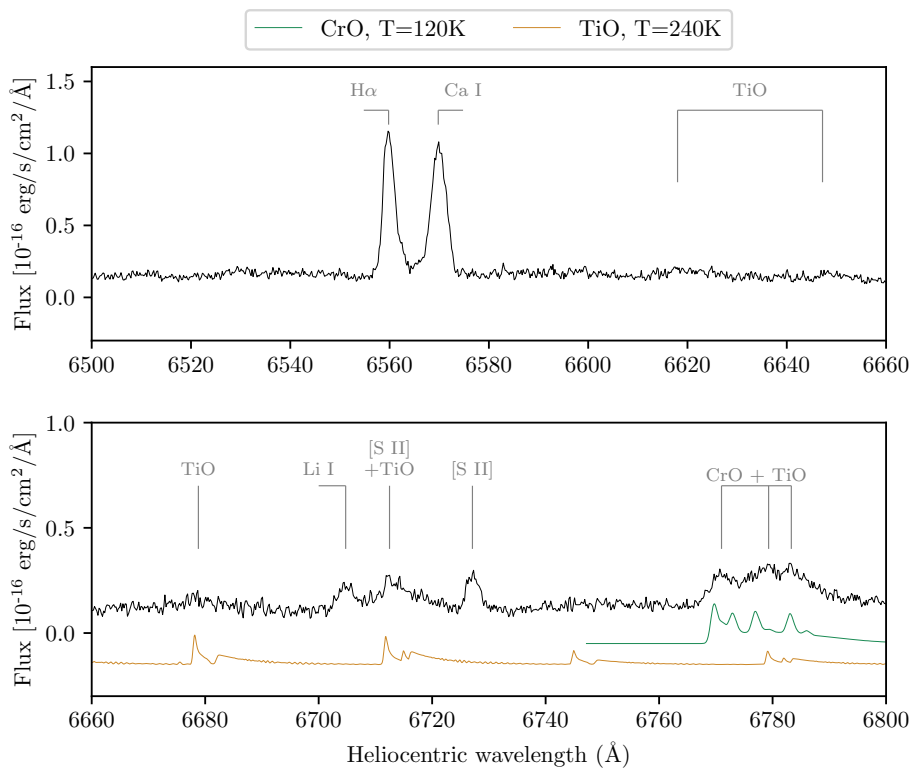}
    \caption{Spectra of V1309\,Sco from 2016 near the lithium and calcium resonance lines. Neighboring features are identified. To better illustrate the contribution of TiO and CrO emission, simple simulations of their emission are shown in orange and green, respectively. The simulations were performed as in \cite{Kamiv1309}. Spectra were dereddened, but were not corrected for the spectroscopic baseline (which is not the true continuum of V1309\,Sco).\\}
    \label{fig-v1309-ola}
\end{figure*}

\begin{figure}
    \centering
    \includegraphics[width=\columnwidth]{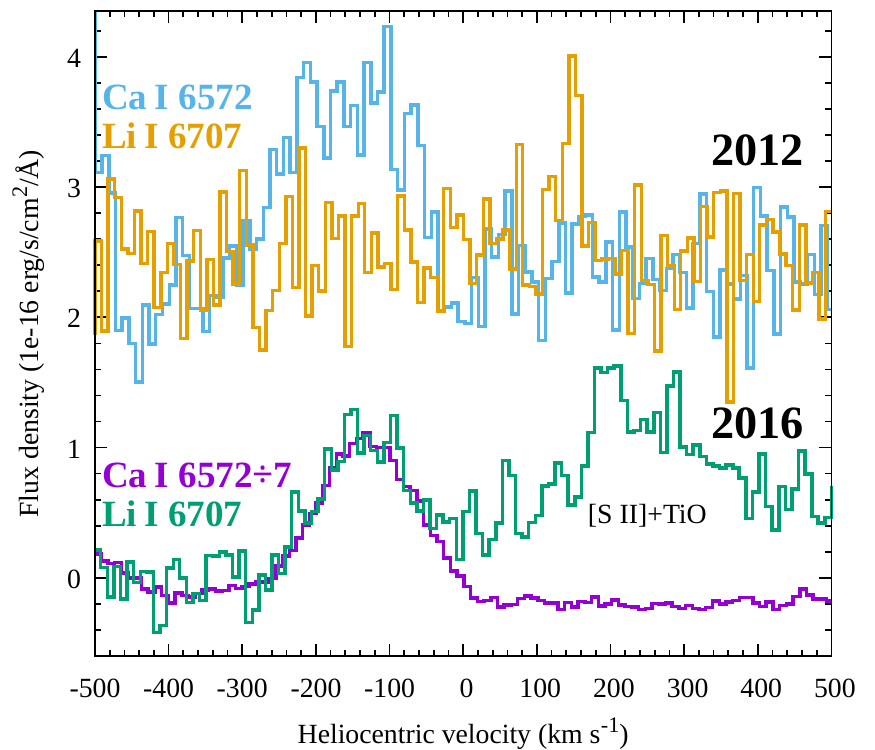}
    \caption{Profiles of the Ca\,I and Li\,I resonance lines in X-shooter observations of V1309\,Sco from 2012 (upper) and 2016 (lower). The Ca\,I flux in 2016 was divided by 7. In 2016, at velocities higher than about 0\,\kms, the Li\,I line partially overlaps with a blend of TiO and [S\,II].}
    \label{fig-v1309-profiles}
\end{figure}

Lithium features have not been reported in the spectra of V1309\,Sco before this work. We find the $\lambda$6707 feature in emission in an X-shooter spectrum from May--July 2016. The spectrum was reported by \cite{KamiSubmm}. Parts of the spectrum covering the lithium and calcium lines of our interest are shown in Fig.\,\ref{fig-v1309-ola}. The Ca\,I line is very close to H$\alpha$ with both lines having nearly equal strength, but the Ca\,I line is relatively clear from contamination by H$\alpha$. The Li\,I line partially overlaps with a broad blend composed of the TiO $\gamma$ 1--0 band and a line of [S\,II] (at a rest wavelength of 6716.44\,\AA). As shown in Fig.\,\ref{fig-v1309-profiles}, the lithium and calcium lines have almost the same profiles, but the exact appearance relies on an arbitrary definition of the local baselines. The measured line fluxes of $F_{\rm CaI}$=33.2$\pm$0.7 and $F_{\rm LiI}$=4.8$\pm$0.4 10$^{-16}$ erg\,s$^{-1}$cm$^{-2}$ yield a calcium to lithium flux ratio of 6.9$\pm$0.6, but uncertainty in this value is larger than indicated owing to uncertain baseline levels for both lines. 

The spectrum of the V1309\,Sco remnant is dominated by emission lines formed through resonant scattering \citep{Chandrasekhar}, whereby absorption of a stellar photon from the ground state is followed by emission to the ground. The emission is thus coupled to the incident light of the star. This kind of spectrum was well studied in V4332\,Sgr, a spectroscopic twin of V1309\,Sco \citep{spectropolarimetry,GornySpec}. Assuming that the lines of Li\,I and Ca\,I are formed in the same volume of gas, the number density ratio of both species can be calculated as: 
$$\frac{n_{\rm LiI}}{n_{\rm CaI}} = \frac{F_{\rm Li}f_{\rm CaI}\lambda_{\rm CaI}^2}{F_{\rm CaI}f_{\rm LiI}\lambda_{\rm LiI}^2} \frac{F^*_{\rm CaI}}{F^*_{\rm LiI}} \exp{-(\tau_{\rm CaI}-\tau_{\rm LiI})}.$$
Here, $F$ and $F^*$ represent the integrated line fluxes and the incident starlight fluxes at the wavelengths of the lines. Since the photospheric spectrum of V1309\,Sco was not observed directly (being attenuated by circumstellar dust since before 2012, see \citealp{Kamiv1309}), the ratio of the intrinsic stellar fluxes is unknown. Nevertheless, due to many analogies between V1309\,Sco and V4332\,Sgr, we can assume that the stellar spectrum is close to a giant with an effective temperature of 3200\,K, as determined for V4332\,Sgr in \cite{Kami2010v4332}. Synthetic spectra of giants at this temperature \citep{marcs} show a flux ratio $F^*_{\rm CaI}/F^*_{\rm LiI}\approx$1.5. However, the incident spectrum that is scattered by the Ca\,I and Li\,I lines is already affected by circumstellar absorption of TiO which greatly increases the ratio. For instance, in the spectrum of V838\,Mon observed in 2005, this ratio of about 4.5 \citep[][their Fig.\,3]{KamiKeck}. We therefore adopt a value of 3.0$\pm$1.5 for V1309\,Sco. 

Although the calcium line is a semi-forbidden transition, it can have a significantly higher optical depth than the seven times weaker line of lithium. Because we do not have a reliable constraint on the size of the emission region, calculating line opacities solely from the observed fluxes is not possible, and thus we ignore the saturation effects, that is, take $\tau_{\rm CaI}$=$\tau_{\rm LiI}$=0. 

We obtain the density ratio $n_{\rm LiI} / n_{\rm CaI}$=2.8($\pm1.7$)$\cdot$10$^{-5}$ and the absolute abundance $A$(Li)=1.8($\pm0.3$). The lithium abundance is a few times higher than in the Sun ($A_{\sun}$(Li)=1.1), but is also a few times lower than in the two other red nova remnants analyzed here. 

\begin{figure}
    \centering
    \includegraphics[width=\columnwidth]{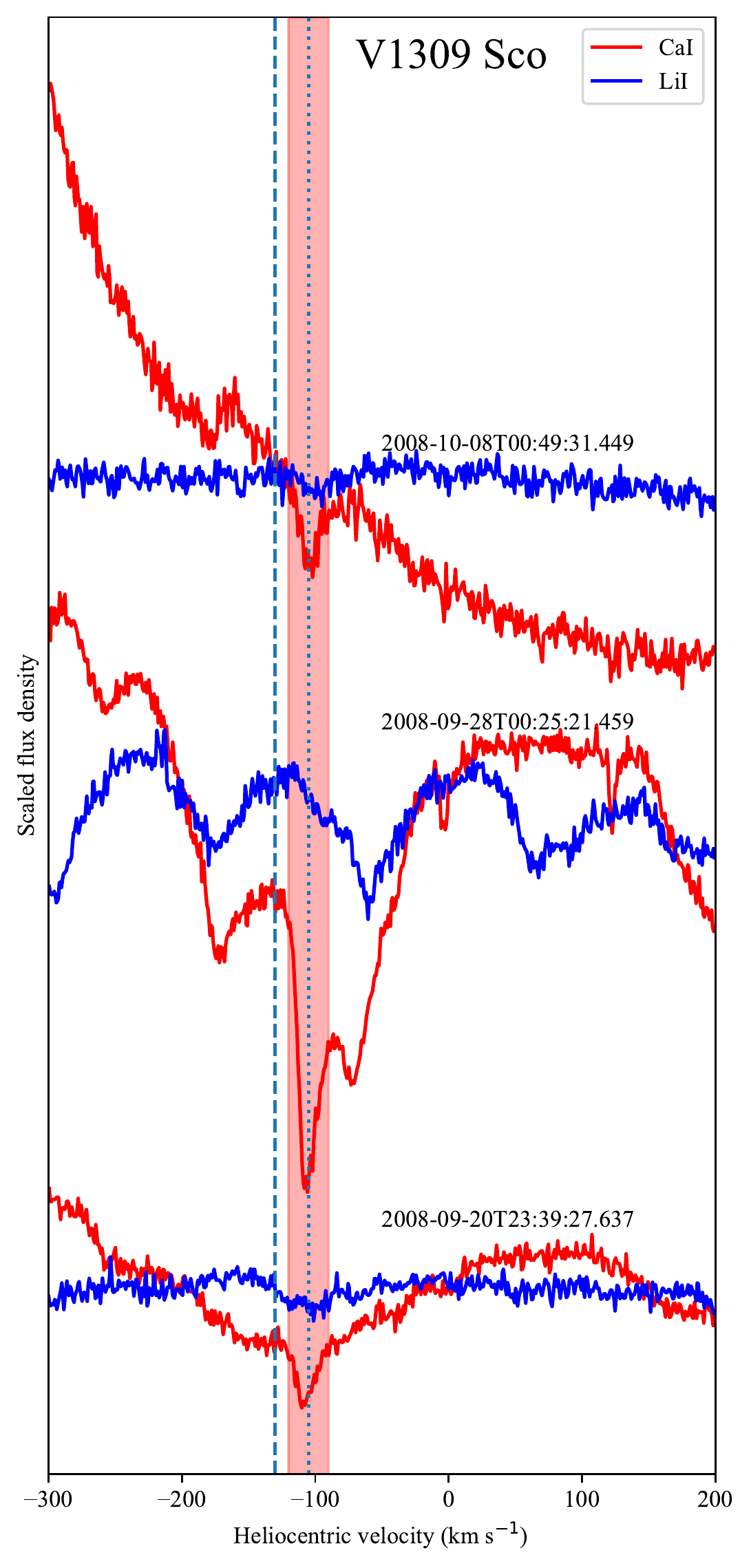}
    \caption{Sample spectra near the Li\,I and Ca\,I resonance lines as observed by UVES in 2009 in V1309\,Sco \citep{Mason}. The observing epochs are indicated as UT times for each pair. Spectra were not normalized and for each pair are in scale. The area highlighted in pink is where strongest Ca\,I absorption is observed and where Li\,I may have a weaker corresponding feature. The vertical dotted line marks the central velocity of this component. The dashed line marks the peak position of the Li\,I and Ca\,I emission lines in the 2016 spectrum. }
    \label{fig-v1309-UVES}
\end{figure}

Lines of both species are centered at --130 ($\pm$10)\,\kms\ and occupy a velocity range typical for most other neutral resonance lines in the late 2016 spectrum of V1309\,Sco, for example the K\,I doublet \citep[cf.][]{Kamiv1309, KamiSubmm}. This velocity is also close to our best constraints on the systemic heliocentric velocity of the object, that is, --110 or --120\,\kms\ \citep{KamiSubmm,MasonShore}. As shown in Fig.\,\ref{fig-v1309-profiles}, the calcium line was also present in a spectrum obtained in 2012 and reported in \cite{Kamiv1309}. However, the lithium line was not detected in 2012, most likely due to a poorer signal-to-noise ratio---the non-detection is consistent with the line ratio we measured for the 2016 features. Even earlier spectra, as those from 2009 presented by \cite{Kamiv1309}, do not show any feature near 6707\,\AA. We also analyzed UVES spectra from \cite{Mason} taken during the outburst of V1309\,Sco, that is, in September--October 2008, but do not find any features at velocities near --130\,\kms\ that could be assigned to the Ca\,I and Li\,I resonance lines. However, multiple spectra in this dataset show broad absorption features centered at $\approx$--105\,\kms\ which can be assigned to Ca\,I and which may have a very faint corresponding feature of lithium. We find these lines in spectra from 18, 20, 28 September 2008, 8 and 20 October 2008 -- but not in spectra from early September 2008. Sample spectra are shown in Fig.\,\ref{fig-v1309-UVES}. The relatively low expansion velocity of the calcium-bearing gas places it within the ``low-velocity component'' of \cite{MasonShore}, which -- according to these authors -- originates in material lost from the V1309\,Sco system before the red nova outburst and which was excited in 2008--2009 by the radiation of the outburst. If the weak features that we have identified indeed belong to Li\,I, the enrichment in lithium may even precede the 2008 eruption of V1309\,Sco, possibly when the progenitor system of subgiants \citep{Stepien} was losing mass. Some solar-mass subgiants are known to have super-solar lithium abundances \citep{lithiumSubgiants}, and thus the progenitors could have been a direct source of the lithium reservoir. However, the spectra from the outburst phase 2008--2009 are very complex, and it is difficult to make unambiguous identifications of the observed features. While it is certain that the remnant displays a lithium enrichment in circumstellar gas in 2016, it is uncertain for dates earlier than 2016. 

If we accept that lithium was only seen in late type spectra, its production can be similar to that discussed for V838\,Mon, that is, through internal production due to enhanced mixing caused by the merger or premerger phases.

\paragraph{V4332 Sgr} V4332\,Sgr is a close analog of V1309\,Sco \citep{Kamiv1309,KamiSubmm}. Its outburst was observed in 1994, that is, twelve years before that of V1309\,Sco. No lithium feature has ever been reported in the spectra of V4332\,Sgr. Spectra obtained during and soon after the 1994 eruption \citep{martini,kimeswenger2006,BanerjeeSpec,TylendaSpec} do not show any conspicuous features that could be assigned to Li\,I $\lambda$6707. However, most of these spectra have very modest spectral resolutions (<10\,000) and signal-to-noise ratios, and thus we cannot definitively claim absence of lithium.  Because V4332\,Sgr is an oxygen-rich object, its strong photospheric and circumstellar bands of TiO $\gamma$ 1--0 hamper any meaningful tests on the presence of Li\,I  $\lambda$6707 in the poor quality spectra. Also, there was a gap of eight years between the earliest spectra obtained in 1994 by \citet{martini} and the next one, i.e. from August 2002 \citep{kimeswengerNA}. The best-quality spectrum of the V4332\,Sgr remnant was obtained with UVES in 2005 \citep{GornySpec} and it does not show any Li features. If the Li\,I line was a temporal feature in V4332\,Sgr, it could have been missed.

\section{Conclusions}\label{final}
Some overabundance of lithium is observed in all the three well-observed red-nova remnants, V838 Mon, V1309\,Sco, and CK\,Vul. The origin of the enhancements is however not clear and, as discussed, may have a different origin in different objects. However, it is always liked to the Cameron-Fowler mechanism that requires deep mixing. Because in some objects there are hints of Li overabundance during the outburst (V838\,Mon) and possibly even before it (V1309\,Sco), some convective configuration may be responsible. Since common envelope evolution appears to be a common feature in these objects, deep convection in that phase is a promising scenario explaining lithium in red novae. With this observational work, we would like to encourage theoretical studies of lithium production through activation of the Cameron-Fowler mechanism in common-envelope systems, merger events, and in merger remnants.

The curious presence of lithium in red novae and their remnants does not exhaust links between lithium production and common-envelope evolution or stellar mergers. Lithium enhancements are often found in non-binary giants with a history of enhanced mass loss and with rapid rotation, including small subsets of RGB stars \citep[e.g.][]{delaReza} and AGB stars \cite[e.g.][]{Reyniers}. A link of some of them to red novae was recently discussed by \cite{Melis}. It also appears now that mergers are the only sensible way to explain J-type and early R-type carbon stars rich in lithium \citep{Izzard,Jstars2020}. Finally, the resonance line of lithium was observed in extragalactic gap transients, including an {\it intermediate-luminosity red transient} AT\,2019abn \citep{Williams2020} and possibly a {\it luminous red nova} AT\,2017jfs \citep[][priv. comm.]{Pastorello2019}, which may as well be powered by mergers. Lithium appears to be an important element for understanding a wide range of systems suspected of common envelope evolution or mergers. 

This study identifies red novae and their remnants as copious lithium producers. However, given their rarity, now estimated to two bright outbursts per decade \citep{Kochanek}, they are unlikely to be key players in the Milky Way lithium budget.

The derivation of circumstellar lithium abundances was rather difficult in red nova remnants. Potential depletion of calcium into dust make our estimates of $A$(Li) upper limits, and further studies of dust formation in these objects may improve on the uncertainties. Radiative transfer modelling of the spectra of these objects in different phases are underway to better address line opacity effects.


\begin{acknowledgements}
We are grateful to: Th. Rauch for kindly sending us spectra from the early outburst of V838\,Mon; P. Martini for providing us with spectra of V4332\,Sgr; A. Pastorello for sharing the spectra of AT\,2017jfs; and Y. Cai for sharing data on AT 2021biy. We thank R. Tylenda for comments on the manuscript and Y. Pavlenko for attempts in modeling the high-excitation lines of lithium in V838\,Mon.

T.K. and I.I. acknowledge funding from grant SONATA BIS no 2018/30/E/ST9/00398 from the Polish National Science Center. 

Based on observations collected at the European Organisation for Astronomical Research in the Southern Hemisphere under ESO programmes 382.D-0152, 088.D-0112, 60.A-9445, 089.D-0041, 097.D-0092, and 281.D-5055. 

This research is based in part on data collected at the Subaru Telescope, which is operated by the National Astronomical Observatory of Japan. We are honored and grateful for the opportunity of observing the Universe from Maunakea, which has the cultural, historical, and natural significance in Hawaii. 

Some data presented herein were obtained at the W. M. Keck Observatory, which is operated as a scientific partnership among the California Institute of Technology, the University of California and the National Aeronautics and Space Administration. The Observatory was made possible by the generous financial support of the W. M. Keck Foundation. The authors wish to recognize and acknowledge the very significant cultural role and reverence that the summit of Maunakea has always had within the indigenous Hawaiian community. We are most fortunate to have the opportunity to conduct observations from this mountain.

Some observations reported in this paper were obtained with the Southern African Large Telescope (SALT). Polish participation in SALT is funded by grant No. MNiSW DIR/WK/2016/07.

Based on observations obtained at the international Gemini Observatory, a program of NSF’s NOIRLab, which is managed by the Association of Universities for Research in Astronomy (AURA) under a cooperative agreement with the National Science Foundation on behalf of the Gemini Observatory partnership: the National Science Foundation (United States), National Research Council (Canada), Agencia Nacional de Investigaci\'{o}n y Desarrollo (Chile), Ministerio de Ciencia, Tecnolog\'{i}a e Innovaci\'{o}n (Argentina), Minist\'{e}rio da Ci\^{e}ncia, Tecnologia, Inova\c{c}\~{o}es e Comunica\c{c}\~{o}es (Brazil), and Korea Astronomy and Space Science Institute (Republic of Korea).

\end{acknowledgements}
\bibliographystyle{aa}
\bibliography{0bib.bib}

\begin{appendix}
\section{Spectral fits}
\begin{figure}
    \centering
    \includegraphics[trim=25 0 60 10, clip, width=\columnwidth]{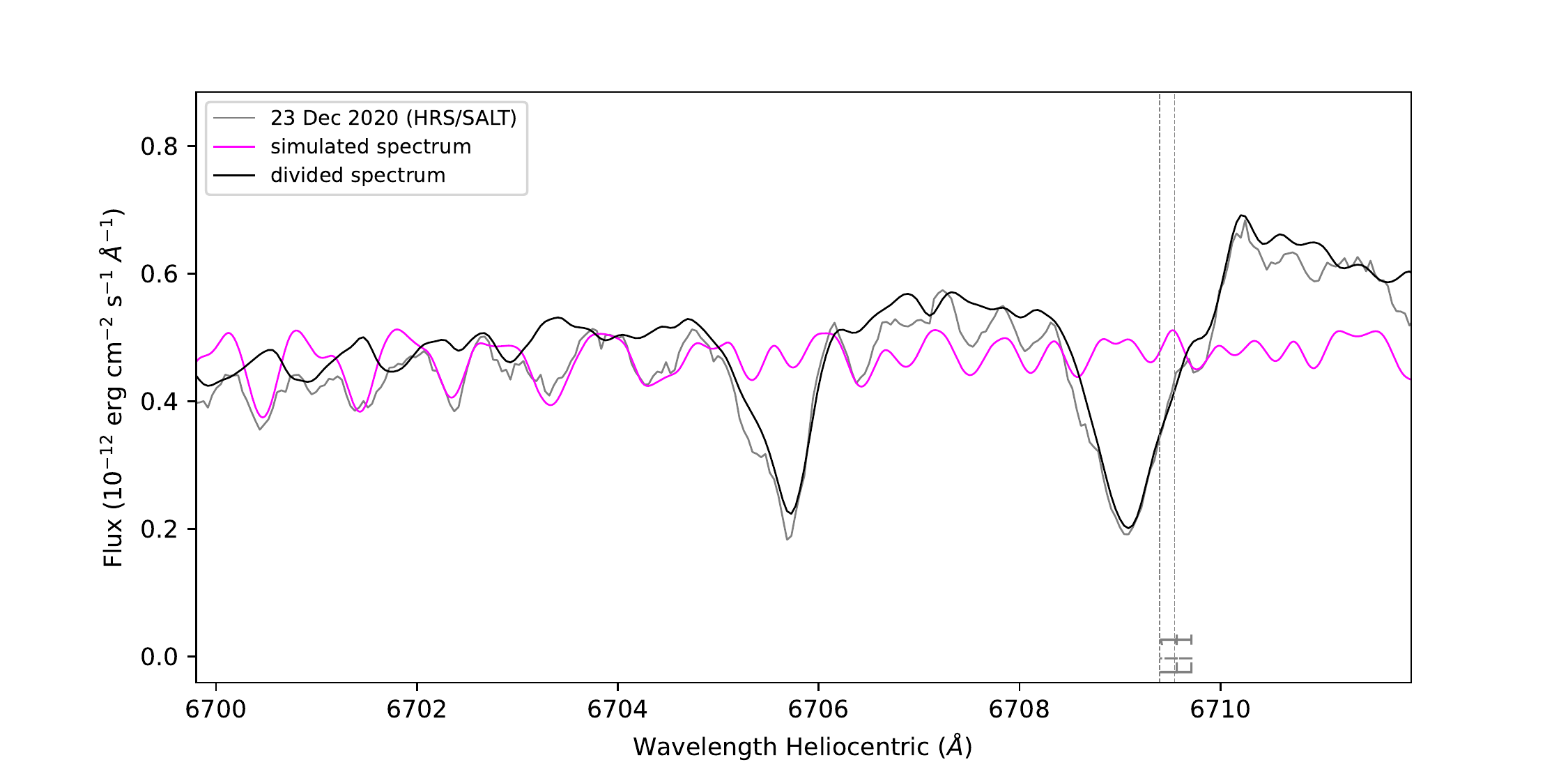}
    \caption{A sample correction of the spectrum near lithium lines by a simulation of TiO bands. The observed spectrum (gray) was obtained with SALT in 2020. The simulation (magenta) included TiO opacity of a cool (800\,K) slab of circumstellar gas. Aside from Li\,I absorption components, most of the spectral features are TiO rotational lines in the $P_2$ ($J_{\rm low}$=24--27) and $Q_2$ ($J_{\rm low}$=36--40) branches. Weak absorption from a couple of Fe\,I lines may be present as well.} 
    \label{fig-v838-fit}
\end{figure}

\begin{figure}
    \centering
    \includegraphics[width=\columnwidth]{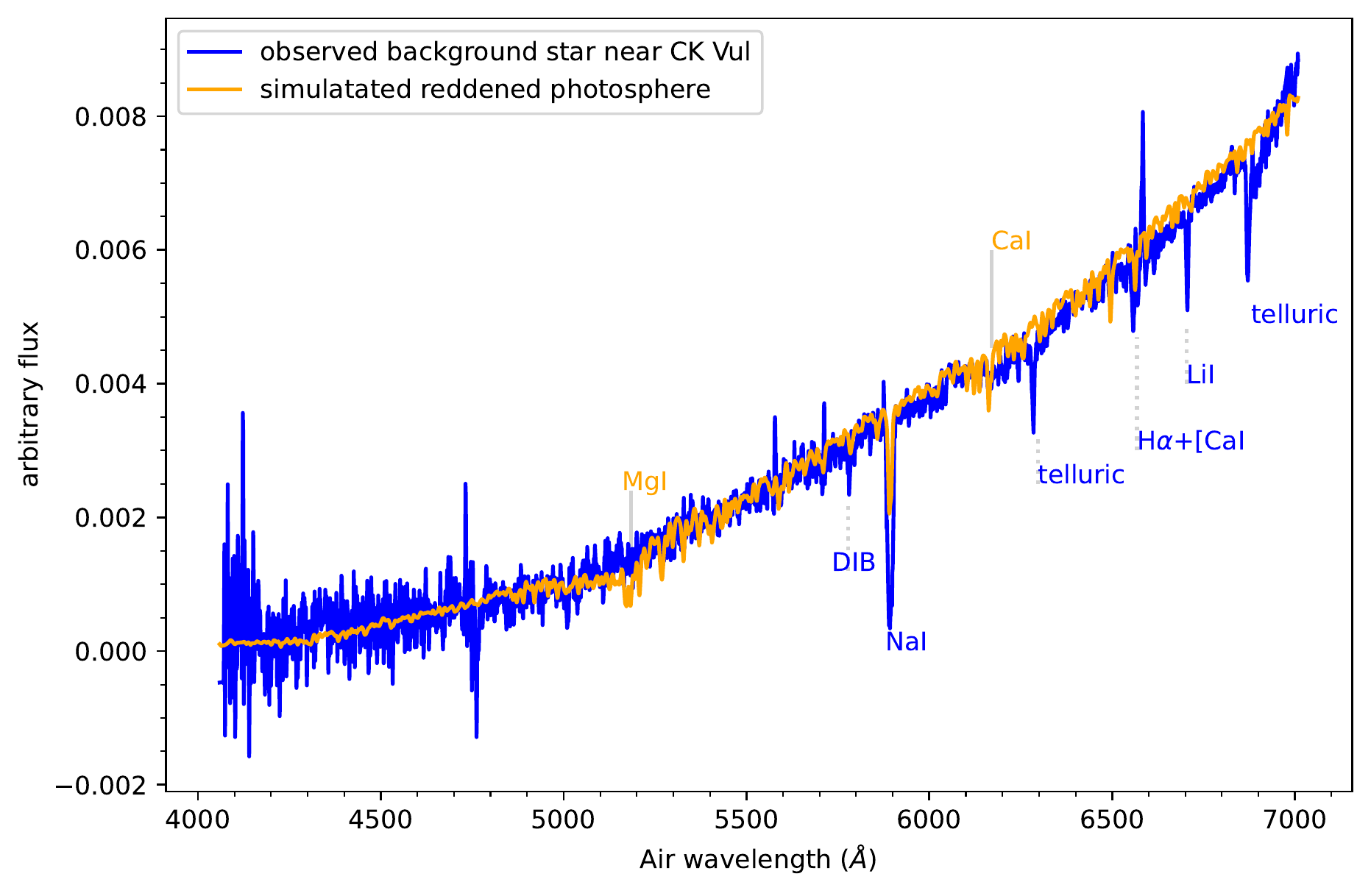}
    \caption{Best fit synthetic spectrum (orange) attempting to reproduce the observed spectrum (blue) of the field star observed behind the outflow of CK\,Vul. The model spectrum was reddened by $A_V$=6\,mag. Main features are identified.}
    \label{fig-ck-fit}
\end{figure}
\end{appendix}

\end{document}